\UseRawInputEncoding
\documentclass[aps,prd,floatfix,superscriptaddress,preprintnumbers,nofootinbib,singlecolumn,longbibliography,letterpaper,natbib,nofootinbib]{revtex4-1}
\pdfoutput=1
\usepackage{amsmath, amssymb, amsfonts, graphicx, mathrsfs, verbatim, float, slashed, bbm, color, xcolor, xspace, enumerate, url, bbding, multirow, outlines,upgreek}
\usepackage[english]{babel}
\usepackage[colorlinks=true,linkcolor=black,citecolor=galacticcenterbubblegum]{hyperref}
\definecolor{galacticcenterbubblegum}{rgb}{0.8,0, 0.8}

\usepackage{graphicx,bm}

\newcommand{\beq}{\begin{equation}}
\newcommand{\bea}{\begin{eqnarray}}
\newcommand{\eeq}{\end{equation}}
\newcommand{\eea}{\end{eqnarray}}
\newcommand{\bal}{\begin{align}}
\newcommand{\eal}{\end{align}}

\usepackage{tikz}
\usetikzlibrary{decorations.pathmorphing,decorations.markings}
\tikzset{
photon/.style={decorate, decoration={snake,amplitude=4pt, segment length=7pt}, draw=black},
particle/.style={draw=black, postaction={decorate}, decoration={markings,mark=at position .5 with {\arrow[draw=black]{>}}}},
antiparticle/.style={draw=black, postaction={decorate}, decoration={markings,mark=at position .5 with {\arrow[draw=black]{<}}}},
gluon/.style={decorate, draw=black, decoration={coil,amplitude=3pt, segment length=4pt}},
higgs/.style={draw=black,dashed,thick },
arrow/.style={draw=black, very thick, postaction={decorate}, decoration={markings,mark=at position 1 with {\arrow[draw=black]{>}}}}
}

\usepackage{latexsym}
\usepackage{subcaption}

\usepackage{outlines}
\usepackage{enumitem}
\setenumerate[1]{label=\Roman*.}
\setenumerate[2]{label=\Alph*.}
\setenumerate[3]{label=\roman*.}
\setenumerate[4]{label=\alph*.}

\newcommand{\cL}{\mathcal{L}}

\newcommand{\cM}{\mathcal{M}}

\newcommand{\DM}{\text{DM}}
\newcommand{\D}{\text{D}}
\newcommand{\R}{{_R}}
\newcommand{\X}{{_X}}

\newcommand{\SU}{SU(2)_{_R}}

\definecolor{darklightsabergreen}{rgb}{0.0, .49, 0.06}

\def\Re{{\cal R \mskip-4mu \lower.1ex \hbox{\it e}\,}}
\def\Im{{\cal I \mskip-5mu \lower.1ex \hbox{\it m}\,}}

\def\tev{\,{\ifmmode\mathrm {TeV}\else TeV\fi}}
\def\gev{\,{\ifmmode\mathrm {GeV}\else GeV\fi}}
\def\mev{\,{\ifmmode\mathrm {MeV}\else MeV\fi}}
\def\to{\rightarrow}

\begin{document}

\vspace*{15mm}
\vspace{1cm}
\title {Light and Feebly Interacting Non-Abelian Vector Dark Matter Produced Through Vector Misalignment }
\preprint{MITP-22-023}
\vspace{1cm}

\author{Fatemeh Elahi}
\email{felahi@uni-mainz.de}
\affiliation{PRISMA$^+$ Cluster of Excellence \& Mainz Institute for Theoretical Physics, \\
Johannes Gutenberg-Universität Mainz, 55099  Mainz, Germany} 
 \author{Sara Khatibi}
 \email{sara.khatibi@ut.ac.ir}
 \affiliation{Department of Physics, University of Tehran, North Karegar Ave., Tehran 14395-547, Iran}

%%%%%%%%%%%%%%%%%%%%%%%%%%%%%%%%%
\vspace*{0.5cm}
\begin{abstract}
\vspace*{0.5cm}
%%%%%%%%%%%%%%%%%%%%%%%%%%%%%%%%%
In this paper, we examine the evolution of light and feebly interacting non-abelian dark gauge bosons dark matter in the early universe. 
We specifically work on a multi-component dark matter scenario with a $SU(2)_\R$ gauge symmetry, spontaneously broken by a scalar $\phi$. 
For sufficiently light $\lesssim O( \rm{MeV})$ and feebly interacting $(g_\R \lesssim 10^{-10})$ gauge bosons,  
$W^3_\R$ and $W^{\pm}_\R \equiv W_\R^1 \pm i W_\R^2$ become the dark matter candidates. 
In this study, the portal between the dark sector and the standard model sector is provided 
by the right-handed electron charged under $SU(2)_\R$. We show that in the region of the parameter 
space where $g_\R \sim 10^{-12}$, the dark gauge boson $W^3_\R$ can be produced efficiently via the freeze-in mechanism, 
however, this mechanism fails to produce $W^{\pm}_\R$ sufficiently. 
For smaller dark gauge coupling, the relic density of three dark matter components can be obtained via the vector misalignment mechanism. 
Vector misalignment has already been discussed in the case of dark photon dark matter. 
In this study, we extend the arguments to non-abelian gauge bosons. 
After discussing the evolution of $W_\R$s in the early universe, we study the model's constraints and show the parameter space's allowed region. 
\end{abstract}
\maketitle

%%%%%%%%%%%%%%%%%INTRODUCTION%%%%%%%%%%%%%%%%%
\section{Introduction}
\label{intro}
%%%%%%%%%%%%%%%%%%%%%%%%%%%%%%%%%
Many observations from inter-galactic to cosmic scales indicate the existence of Dark Matter (DM) 
that corresponds to approximately $25\%$ of the energy budget of the Universe. 
However, the nature and origin of DM are still enigmatic. 
The long sought-after weakly interacting massive particle with $O(\text{GeV})$ mass scale has been 
under extreme experimental scrutiny, and yet, no promising sign has been 
observed~\cite{Cui:2017nnn,Aprile:2018dbl,Akerib:2016vxi}. 
Thus, there has been a slow shift of interest to other DM candidates, especially in the sub-GeV mass regime.  

One of the simplest and well-studied models in this regime is the 
dark photon dark matter~\cite{Pospelov:2008jk,Redondo:2008ec}. 
These models come from extending the Standard Model (SM) 
by a gauge $U(1)_\D$ that is broken either through spontaneous symmetry breaking (SSB) 
or via Stuckleberg mechanism~\cite{Kors:2005uz}, resulting in a massive dark photon. 
The main portal to SM in these studies is the kinetic mixing (e.g, $ \epsilon F'_{\mu \nu} F^{\mu \nu}$). 
In the mass basis, the kinetic mixing term induces a coupling between dark photon and 
the electromagnetic current proportional to $\epsilon$.  
However, for sufficiently light dark photon $(m_{A'} < 2 m_e)$ and small coupling, 
the dark photon may be long-lived and can account for the relic abundance of DM. 

The story of SM $+~U(1)_\D$ is particularly simple because it does not carry any adjoint index,
and thus, the kinetic mixing term is relevant even at low scales.
Extending the SM with a non-abelian symmetry, on the other hand, 
has extra complications because the associated gauge bosons carry an adjoint index. 
In this case, the kinetic mixing operator between $SU(N)_\D$ and $U(1)_Y$ is non-renormalizable and usually ignorable.
The common practice, in such models, is using the Higgs portal or introducing a vector-like fermion 
which is a doublet of the new gauge symmetry~\cite{Belyaev:2022zjx,Belyaev:2022shr} 
or charging some SM particles under $SU(N)_\D$. 
In this paper, we will use the latter, and charge (a subset of) right-handed particles 
under an extra $SU(2)$ (refer to as $SU(2)_\R$). 
Since the phenomenology of electrons is relatively the richest, 
we consider the right-handed electron to be charged under the $SU(2)_\R$ as a portal between the dark and the SM sectors.

In this paper, we consider a multi-component DM scenario with a $SU(2)_\R$ gauge symmetry, 
spontaneously broken by a scalar $\phi$ which is a doublet under the $SU(2)_\R$.
In contrast to the $U(1)$, in a non-abelian gauge symmetry, we cannot use the Stuckelberg mechanism 
for having the massive gauge bosons and so we should employ the Higgs mechanism.
Thus after the SSB, we have three massive dark gauge bosons, 
$W_\R^3$ and $W_\R^\pm \equiv W_\R^1 \pm i W_\R^2$.
We show that in the region of parameter space in which we are interested (light mass and small coupling), these three gauge bosons 
are good DM candidates. 
Due to the presence of $\phi$ in the scalar potential, the Higgs portal is inevitable.
To restrain the influence of the Higgs portal, we study the limit where $v_\phi \gg v_h$. 
Since $m_{_{W_\R}} \propto g_\R v_\phi$ and we are interested in $m_{_{W_\R}}$ 
in the sub-GeV scale, we are led to consider very small gauge couplings.
%This region of the parameter space is further motivated due to the recent 
%XENON1T-excess in the electron recoil analysis~\cite{Aprile:2020tmw}. 
%This excess is still not substantiated. Nonetheless, in case it is a sign of new physics, 
%it has already been demonstrated that dark photon dark matter with 
%mass $\sim 2 \ \text{keV}$ and $\epsilon \simeq 10^{-16}$ can explain 
%this excess~\cite{Alonso-Alvarez:2020cdv,An:2020bxd,Okada:2020evk,Lindner:2020kko}. 
%We also point out that this benchmark in our model can explain the XENON1T excess as well.
 
The production of dark gauge bosons with such feeble couplings in the early universe is non-trivial. 
The small coupling prevents the dark gauge bosons from staying in thermal equilibrium.
The canonical solutions are either freeze-in 
mechanism~\cite{Hall:2009bx,Elahi:2014fsa,Okada:2020evk, Delaunay:2020vdb, Okada:2020cue}, 
inflationary fluctuations~\cite{Graham:2015rva,Alonso-Alvarez:2018tus,Ema:2019yrd,Ahmed:2020fhc, Arvanitaki:2021qlj}, 
vector misalignment~\cite{Nelson:2011sf,Arias:2012az,Alonso-Alvarez:2019ixv,
Nakayama:2019rhg,Nakayama:2020rka}, 
or the efficient transfer of energy from an axion (or axion-like particle) 
to dark gauge boson~\cite{Agrawal:2018vin,Co:2018lka,Dror:2018pdh}.
Most of these mechanisms have been studied in great detail for the case of a dark photon, 
but have not been employed for the case of non-abelian dark gauge bosons, as far as we know. 
In this paper, we will discuss the region of the parameter space ($g_\R \sim 10^{-12}$) that 
freeze-in can explain the relic abundance for $W_\R^3$ DM. 
However, the freeze-in mechanism cannot produce sufficient relic density for $W_\R^\pm$.
For even smaller couplings, we extend the analysis on vector misalignment to $SU(2)_\R$ dark gauge boson dark matter,
and we show that all three components of dark gauge boson can be produced via vector misalignment.
Similar to dark photon dark matter production via misalignment mechanism, the density of $W_\R$s depletes like $a^{-2}$ during inflation. 
However, there are various solutions to avoid this depletion 
(e.g., adding a non-minimal coupling between $W_\R$s and the Ricci Scalar~\cite{Arias:2012az,Alonso-Alvarez:2019ixv}, 
or introducing coupling between the Inflaton and $W_\R$s~\cite{ Nakayama:2019rhg,Nakayama:2020rka}). 
In the following sections, we discuss the relative advantages and disadvantages of some of these solutions.
In short, the main phenomenological difference between these solutions 
is the constraint on the Hubble scale in the inflation era ($H_I$), which is currently not measured.  

The organization of the paper is as follows. In Section~\ref{sec:model}, 
we explain the model and the new degrees of freedom. 
Section~\ref{sec:Pro} is dedicated to various means of $W_\R$s production, 
including freeze-in (Subsection~\ref{sec:FI}), and vector misalignment (Subsection~\ref{sec:Misalignment}). 
In section~\ref{sec:Pheno}, we discuss the phenomenological implication of the model, 
and the concluding remarks are presented in section~\ref{sec:Sum}. 

 %%%%%%%%%%%%%%%MODEL %%%%%%%%%%%%%%%%%%%
\section{DM Model}
\label{sec:model}
%%%%%%%%%%%%%%%%%%%%%%%%%%%%%%%%%%%%%%
We consider a multi-component DM scenario, where the SM symmetry group is enlarged by a  gauge $\SU$ symmetry, 
that is spontaneously broken by $\phi$ which is a doublet of $SU(2)_\R$, and singlet of SM gauge groups.
The subscript $R$ indicates that (a subset of) right-handed particles are charged under $SU(2)_\R$. 
Since the phenomenology of electrons is relatively the richest, 
we narrow our attention to the case where only the right-handed electron is a doublet of $SU(2)_\R$: $E_\R =  \left( e' ~~~ e\right)^T_\R $. 
The SM electron is shown by $e$ and the $e'$ is a new fermion with the same quantum numbers as the right-handed electron. 
We charge the right-handed electron under $SU(2)_\R$ to introduce a portal between SM and dark sectors.
Since $E_\R$ has $U(1)_Y$ charge as well, our model is anomalous. 
Ref.~\cite{Elahi:2019jeo} discusses the solution to the anomaly problem in detail. 
Excluding the particles needed in the UV to make the model anomaly free, the Lagrangian of the model, 
governing the phenomenology of the model, becomes the following:

\begin{align}
\label{eq:lag}
\cL&= \cL_{_{\text{SM}}}+ \frac{1}{4} W^{ a}_{_{R}\mu\nu} W^{ a \mu\nu}_{_R}+  
(D_{\mu}\phi)^{\dagger}(D^{\mu}\phi)- \mu_{_{\phi}}^2 \phi^{\dagger}\phi  
+ \lambda_{_{\phi}} (\phi^{\dagger}\phi)^2 \\ \nonumber
&+\bar{E_{_R}}(\imath D\!\!\!/ )E_{_R} + \left( \frac{y_{e}}{\Lambda_1}(\bar{L}H)(\phi^{\dagger}E_{\R})
+ h.c.\right)\\ \nonumber
&+ \xi_{H \phi} (H^{\dagger}H)(\phi^{\dagger}\phi)+ \epsilon_{_W} \frac{(\phi^\dagger \tau^a \phi)}{\Lambda_2^2}   
W^{a}_{\R\mu \nu}B^{\mu \nu},
\end{align} 
where $W^{a }_{\R \mu\nu}$ is the field tensor of $SU(2)_\R$ and $B_{\mu \nu}$ is that of hypercharge.  
The covariant derivative is defined as 
$ D_\mu  = \partial_\mu + i g_{\R}  \frac{\tau^a}{2} W_{_{R} \mu}^a + i Q_Y g_{_Y} B_{_\mu} $, 
where $g_{\R}$ is the new gauge coupling and $g_{_Y}$ is the hypercharge. 
Note that $\phi$ is not charged under $U(1)_Y$, and thus $Q_Y^\phi = 0$, while $Q_Y^{E_\R} = -1$. 
After $\SU$ SSB, the three dark gauge bosons acquire the same mass,
\begin{equation*}
m_{_{W_\R}} = \frac{g_\R v_\phi}{2}, 
\end{equation*}
where the $v_\phi$ is the vacuum expectation value (vev) of the $\phi$. Similarly, 
the mass of $\phi$ becomes $m_\phi =\sqrt{2 \lambda_\phi} v_\phi$.

\begin{table}[!t]
\begin{center}
\begin{tabular}{|ccccc|}
\hline 
 && $\hspace{0.25 in}SU(2)_{_R}\hspace{0.25 in}$ && $\hspace{0.25 in}U(1)_X\hspace{0.25 in}$   \\
\hline
$W_{_{R}}$ && 3 && $0$   \\

$\phi$ && 2 && $1/2$   \\

$E_{_R}$ && 2 && $1/2$\\
\hline
\end{tabular}
$ \xrightarrow{\text{SSB}}$
\begin{tabular}{ |ccc|}
\hline 
 &&  $\hspace{0.45 in} U(1)_\D \hspace{0.45 in}$\\ 
 \hline
$e'$ && $1$\\
$W_\R^\pm$  &&$\pm1$\\
$W_\R^3$ &&0 \\
\hline 
\end{tabular}
\end{center}
\caption{The charges of the new particles under the $ \SU \times U(1)_X   \xrightarrow{\text{SSB}} U(1)_D$ symmetries.
Due to the residual $U(1)_D$, in the limit where $m_{_{W_\R}} \ll m_{e'}$, $W^\pm_\R$ are stable DM candidates.}
\label{tab:Particle}
\end{table} 
 
In this setup, because $e_\R$ is charged under $\SU$, its Yukawa coupling is 
modified as the second term in the second line of Eq.~\ref{eq:lag}.
The SM Yukawa coupling is induced, after $\phi$ gains a non-zero vev: $y_{e}^{\text{SM}}=y_{e} v_{\phi}/\Lambda_1$. 
The mass of $e'$, on the other hand, comes from the fermions in the UV that 
are added to cancel the gauge anomaly (for more details, 
see Ref.~\cite{Elahi:2019jeo}); thus, $m_{e'}  \sim O\left(v_\phi\right) \gg m_e$. 
The third line of Eq.~\ref{eq:lag} shows the Higgs portal, 
and the non-renormalizable kinetic mixing term between $\SU$ and $U(1)_Y$. 
We are interested in the limit where $v_\phi \gg v_h$, so that the mixing angle between the two scalars is close to 0:
\begin{equation*}
\alpha \equiv \frac{1}{2} \text{tan}^{-1}\frac{\xi_{H \phi} v_h v_\phi}{v_h^2 \lambda_H - v_\phi^2 \lambda_\phi} \simeq 0,
\end{equation*}
closing the Higgs portal. 
One may wonder that in the limit of large $v_\phi$, the kinetic mixing term becomes important. 
The main consequence of this term is the mixing between $W_\R^3$ and $B$, 
which induces a new contribution to $W_\R^3 ee$ coupling:
$g_\R + \epsilon_{_W} g_{_Y} \frac{v_\phi^2}{\Lambda_2^2}.$
Given that the ``natural" value of $\epsilon_W$ is about $g_\R g_Y/(16 \pi^2) < g_\R$, and $v_\phi/\Lambda_2 <1$, 
we expect $\epsilon_{_W} g_{_Y} \frac{v_\phi^2}{\Lambda_2^2} \ll g_\R$. 
Hence, for the rest of the paper, we will neglect this term, unless stated otherwise.  

Studying Eq.~\ref{eq:lag}, we notice that there is an accidental global $U(1)_\X$ symmetry with charges 
shown in Table~\ref{tab:Particle}. Since $\phi$ has a non-zero charge under $U(1)_\X$, 
once $\phi$ gets a vev, $SU(2)_\R \times U(1)_\X$ is spontaneously broken to a residual global $U(1)_D$.
The charges of particles after SSB follows the relation $(I_3)_\R + X$, 
where the $X$ is $U(1)_X$ charge. 
The existence of this symmetry at the low energy guarantees the stability of $W^\pm_\R$ in the limit where $m_{_{W_\R}} < m_{e'}$ 
(since in this limit the $W^\pm_\R \rightarrow e' e$ decay channel is kinematically closed).  
Furthermore, for the light mass limit $m_{_{W_\R}} < 2 m_e$, 
the $W_\R^3$ is not stable and can decay to $3 \gamma$ or $\nu \bar{\nu}$  (through a loop of electrons)
\footnote{The two-photon decay channel is forbidden by the Yang theorem~\cite{Yang:1950rg}}, 
however, for sufficiently small coupling, it can be a long-lived DM candidate. As a result, in this model, 
we have two stable DM particles $(W^\pm_\R)$ and a long-lived DM particle $(W_\R^3)$.
In the following, we will discuss the production of dark gauge bosons DM as well as the phenomenological constraints on this model.

%%%%%%%%%%%%%%%%DMcandidates%%%%%%%%%%%%%%%%%%
\section{DM Production}
\label{sec:Pro}
%%%%%%%%%%%%%%%%DMcandidates%%%%%%%%%%%%%%%%%% 

In this section, we discuss the production of the non-abelian dark gauge bosons 
in the early universe and their relic density at present. 
We are interested in the regime where dark gauge bosons are light (e.g, $m_{_{W_\R}}\sim (\text{eV- MeV})$) 
and feebly interacting. The gauge coupling, in this regime, is small enough 
that prevents dark gauge bosons from reaching thermal equilibrium in the early universe. 
So they cannot be produced through thermal processes. 
Instead, different non-thermal production mechanisms can produce light DM abundance.
The freeze-in mechanism is one of the non-thermal procedures that can produce the dark gauge 
bosons~\cite{Okada:2020evk, Delaunay:2020vdb, Okada:2020cue}.
Moreover, the light dark gauge bosons can be produced during inflation by the vector misalignment 
mechanism~\cite{Nelson:2011sf,Arias:2012az,Alonso-Alvarez:2019ixv,Nakayama:2019rhg,Nakayama:2020rka}. 

In the following, we discuss how and under what conditions the freeze-in and vector misalignment mechanisms 
can produce the correct relic density for the dark gauge bosons in our non-abelian DM scenario.

%%%%%%%%%%%%%%%%%%%%%%%%%%%%%%%%
\subsection{Freeze-in}
\label{sec:FI}
%%%%%%%%%%%%%%%%%%%%%%%%%%%%%%%%%
Assuming dark gauge bosons start with negligible abundance at the reheat Temperature ($T_{\rm{RH}}$), 
they can slowly get produced through their feeble interactions with right-handed $e$ and $e'$. 
This mechanism is known as freeze-in~\cite{Hall:2009bx}. 

To produce the $W_\R^3$, among the possible processes, the contribution from $e e \to \gamma W_\R^3$ 
is dominant \footnote{Interested readers are encouraged to see Appendix~\ref{app:FIsmall} 
for a detailed discussion on the contribution of other potential processes.}. 
The $W^{3}_\R$ density within the freeze-in framework can be obtained 
from the Boltzmann equation~\cite{ Hall:2009bx,Okada:2020evk,Kolb:1990vq}:
\beq
\dot n_{W_\R^3}+ 3 H n_{W_\R^3} = \frac{ T}{256 \pi^6} \int_{m_{_{W_\R}}^2}^\infty ds\  d\Omega\  p_{ee} \ p_{\gamma _{ W_\R^3}}|\mathcal{M}|^2_{ee \to \gamma _{W_\R^3}}K_1(\sqrt{s}/T)/\sqrt{s},
\label{eq:BE}
\eeq
where $p_{ij}\equiv \sqrt{\frac{(s- (m_i + m_j)^2)(s- (m_i - m_j)^2)}{4s}}$, and 
\beq
|\mathcal{M}|^2_{ee \to \gamma _{W_\R^3}} = \frac{8 \pi \alpha g_\R^2 (m_e^4 + 3 m_{_{W_\R}}^2 m_e^2 -2 u m_e^2+ tu)}{(t - m_e^2)^2},
\eeq 
with $s,t,u$ being the ordinary Mandelstam variables. 
After doing the integrals in Eq.~\ref{eq:BE} and changing the 
variable to $Y_{_{W_\R^3}} \equiv n_{W_\R^3}/s$, we get 
\beq
\frac{dY_{_{W_\R^3}}}{dT} \simeq - \frac{32 \pi \alpha g_\R^2}{s(T)H(T)}\left[1+ \text{log}\left(\frac{m_e}{T}\right)\right] T^2 m_{_{W_\R}} K_1\left(\frac{m_{_{W_\R}}}{T}\right), 
\label{eq:dYdT2}
\eeq
in the limit $m_{_{W_\R}} < m_e \ll E_{\text{cm}}$. The Hubble parameter 
and the entropy density of the universe  are shown by $H(T)$ and $s(T)$, respectively, 
and they are given by:
\begin{equation*}
H(T)=1.66 \sqrt{g_{\star}} \frac{T^{2}}{M_{\mathrm{pl}}},~~~~s(T)=\frac{2 \pi^{2}}{45} g_{\star} T^3.
\end{equation*}
where $g_{\star}= 106.75$ is the effective relativistic degrees of freedom 
and $M_{\mathrm{pl}}=1.2 \times 10^{19}$ GeV is the non-reduced Planck mass.
By integrating Eq.~\ref{eq:dYdT2}, with respect to temperature $(T \in [m_{_{W_\R}}, \infty])$, 
the yield of the $W_\R^3$ is obtained~\cite{Hall:2009bx,Kolb:1990vq}:
\beq
Y^0_{_{W^3_{_R}}} \simeq \frac{ 360 \ \alpha \ g_\R^2 M_{\mathrm{pl}}}{1.66 \times g_\star^{3/2} 
m_{_{W_\R}}} \left[0.73 +   \text{log}\left(\frac{m_e}{m_{_{W_\R}}}\right)\right].
\eeq
The relic abundance of $W_\R^3$ produced from $e e \to W_\R^3 \gamma$ is 
\beq
\Omega_{_{W_\R^3}} h^2 =\frac{m_{_{W_\R}} Y^0_{_{W^3_{_R}}}  s_{0} }{\rho_{c}/h^{2} } 
\simeq 10^{24} g_\R^2  \left[0.73 +   \text{log}\left(\frac{m_e}{m_{_{W_\R}}}\right)\right],
\label{eq:relic}
\eeq
where $s_{0} = 2890/\rm{cm}^3$ is the entropy density of the present time 
and $\rho_{c}/h^2=1.05 \times 10^{-5} \ \rm{GeV}/\rm{cm}^3$ is the critical 
density over the square of the reduced Hubble constant.
The Planck collaboration reported $0.119$ value for DM relic density~\cite{Ade:2015xua}. 
As Eq.~\ref{eq:relic} shows, the freeze-in mechanism can explain the relic abundance 
for the ${W_\R^3}$, if $g_\R \simeq 10^{-12}$. However, as we have shown in Appendix~\ref{app:FIsmall},
the freeze-in scenario cannot produce sufficient $W^\pm_\R$ DM in the early universe.

For smaller gauge coupling, we need another non-thermal DM production mechanism.
In the next subsection, we explain how the vector misalignment mechanism can produce 
all three components of the dark gauge bosons in our non-abelian scenario. 

%%%%%%%%%%%%%%%%%%%%%%%%%%%%%%
\subsection{Misalignment}
\label{sec:Misalignment}
%%%%%%%%%%%%%%%%%%%%%%%%%%%%%%

The misalignment is known in the context of scalar/pseudo-scalar 
(like axion and axion-like particles)~\cite{Preskill:1982cy, Abbott:1982af, Dine:1982ah}. 
In the early universe when the Hubble parameter, $H$, is much larger than the mass of a spatially homogeneous scalar, 
the initial value of the scalar takes a random non-zero value. At the transition point, 
when the $H$ is in order of the mass of the scalar, the scalar field starts to oscillate. 
In the late universe when the $H$ is much smaller than the mass of the scalar, 
the field oscillates around the minimum of the potential and its energy density 
is proportional to $a^{-3}$ ($a$ is the scale factor). Thus, it redshifts like non-relativistic matter.

Originally,  Ref.~\cite{Nelson:2011sf} showed that the same production mechanism can apply for an abelian 
light vector DM in a  Friedmann-Robertson-Walker (FRW) background. 
However, at early times (during inflation), the energy density of the vector field is 
proportional to the $a^{-2}$ and the vector field dilutes with expansion and ends up with negligible abundance.
Therefore, multiple solutions were proposed to solve this problem, 
one of them being a new direct coupling between the gauge boson and the curvature scalar~\cite{Arias:2012az, Alonso-Alvarez:2019ixv}.

The works of Ref.~\cite{Arias:2012az,Alonso-Alvarez:2019ixv} are in the context of a dark photon. 
In this paper, we want to extend this mechanism to non-abelian gauge bosons. 
The action of the non-abelian vector field, with non-minimal coupling to the Ricci scalar, is given by:
\beq
S=\int \mathrm{d}^{4} x \sqrt{-g}\left(\frac{1}{2}\left(M_{\mathrm{pl}}^{2}+
\frac{\kappa}{6} W_{_{R} \mu}^b W_{_{R} }^{b \mu}\right) R-
\frac{1}{4} W^{b}_{_{R}\mu\nu} W^{ b \mu\nu}_{_R}-
\frac{1}{2} m_{_{W_\R}}^{2} W_{_{R} \mu}^b W_{_{R} }^{b \mu}\right),
\label{Eq:action}
\eeq
where $R$ is the Ricci scalar, and $\kappa$ is the coupling between the gauge bosons and the Ricci scalar. 
The adjoint index is shown by $b=1,2,3$, and it means that the action can be written for all three components of dark gauge bosons.
The last term in the above action is the mass term for the non-abelian dark gauge bosons that, as explained before, 
comes from the Higgs mechanism. The FRW metric is as follows:
\beq
g=\operatorname{diag}\left(-1, a^{2}(t), a^{2}(t), a^{2}(t)\right),
\eeq
where $a(t)$ is the scale factor.
As shown in the appendix~\ref{app:EOM}, the equations of motion (EOM) for the homogeneous physical field 
($\mathcal{W}^{ b}_{i} \equiv W_{_{R} i}^b/ a$) from the action (Eq.\ref{Eq:action}) can be obtained as follows, 
\beq
\ddot{\mathcal{W}^{ b}_{i}}+3 H \dot{\mathcal{W}^{ b}_{i}}+\left(m_{_{W_\R}}^{2}+
\frac{1-\kappa}{6} R\right) \mathcal{W}^{ b}_{i}=0,
\label{EOM}
\eeq	
and the time component $W_{_{R} 0}^b=0$. 
Therefore, the EOM for the spatial component of a non-abelian vector field is the same as the harmonic 
oscillator with Hubble damping term.
In other words, for $\kappa=1$ the EOM of the vector 
is the same as the EOM of the scalar with minimal coupling to gravity. 
This way, the problem of scaling the dark vector is solved. 
However, it suffers from some other issues (e.g., isocurvature bounds and ghost instability in the theory), 
which we will discuss in more detail later in this section. 
Leaving these issues to a more thorough model building, 
the usual procedure of the misalignment mechanism can be applied for the light non-abelian vector case.

From the above equation, we can see the non-minimal coupling to gravity in the action causes 
the effective mass in the EOM for the non-abelian vector fields. According to the time evolution of the Ricci scalar, the 
effective mass changes, particularly manifests itself in the inflationary era.
Since our non-abelian gauge bosons acquire the mass from the Higgs mechanism,
during inflation, the bare mass of the vector fields is zero.
In the middle of the radiation-dominant era, the SSB happens and the non-abelian vector fields get mass.  
In the appendix~\ref{app:Evolution}, the EOM for the homogeneous non-abelian vector field is solved
for the inflation and post-inflationary era with details. We find that the field value at the end of inflation is as follows:
\beq
\mathcal{W}^{ b}_{i,e} \simeq \mathcal{W}^{ b}_{i,s}  \mathrm{e}^{-\frac{1}{2} \beta_{-} N_{tot}},
\label{Eq:endinflation}
\eeq
where the $\mathcal{W}^{ b}_{i,s}$ is the initial value of the field at the start of inflation, 
the $N_{tot}$ indicates the total number of e-folds of inflation, and $\beta_{-}= 3 - \sqrt{1+8 \kappa}$. 
After inflation and before the SSB happens, the solution of the EOM is $\mathcal{W}^{ b}_{i}(t) \simeq \mathcal{W}^{ b}_{i,e}$.
After the SSB and at the point where $m_{_{W_\R}} \sim \frac{3}{2} H_{\star}$ 
(where $H_{\star}$ is the Hubble parameter at the DM production time), the evolution of the field transits
to an oscillation mode. 
Finally at late times when $m_{_{W_\R}} \gg H $, 
the vector field oscillates around the minimum of the potential and the solution of the EOM is as follows:
\beq
\mathcal{W}^{ b}_{i}(t)= \mathcal{W}^{ b}_{i,e} (\frac{a_{\star}}{a})^{\frac{3}{2}} \cos(m_{_{W_\R}} (t-t_{\star})),
\eeq
where $a_{\star}$ is the scalar factor at the DM production time ($t_{\star}$).
The energy-momentum tensor can be found from the following:
 \beq
 T_{\mu \nu}=\frac{-2}{\sqrt{-g}} \frac{\delta(\sqrt{-g} \mathcal{L})}{\delta g^{\mu \nu}}.
 \eeq
The energy density of the homogeneous non-abelian vector field is as 
follows~\cite{Arias:2012az,Alonso-Alvarez:2019ixv,Golovnev:2008cf}:
\beq
\rho_{_{\mathcal{W}^b}}=T_{00}=\frac{1}{2}\left(\dot{\mathcal{W}^{ b}_{i}} \dot{\mathcal{W}^{ b}_{i}}+m_{_{W_\R}}^{2} \mathcal{W}^{ b}_{i} \mathcal{W}^{ b}_{i}+(1-\kappa) H^{2} \mathcal{W}^{ b}_{i} \mathcal{W}^{ b}_{i}+2(1-\kappa) H \dot{\mathcal{W}^{ b}_{i}} \mathcal{W}^{ b}_{i}\right).
\eeq 
Since in the late time $H \ll m_{_{W_\R}}$, the energy density approximately becomes,
\beq
\rho_{_{\mathcal{W}^b}} \simeq \frac{1}{2} m_{_{W_\R}}^2   (\mathcal{W}^{ b}_{i,e})^2 \left(\frac{a_{\star}}{a}\right)^3.
\eeq
The energy density of the vector field in this region is inversely proportional to the $a^{3}$ (space volume element) 
and thus redshifts like cold matter. 
From the conservation of comoving entropy ($S=s a^3=\frac{2\pi}{45}g_{\star S}(T) T^3 a^3$) we have,
\beq
\left(\frac{a_{\star}}{ a}\right)^{3}=\frac{g_{\star S}(T_0)}{g_{\star S}(T_{\star})} \left(\frac{T_0}{T_{\star}}\right)^3.
\eeq

As a result, the current energy density of the non-abelian vector field becomes,
\beq
\rho^0_{_{\mathcal{W}^b}} \simeq 6.5 ~\left(\frac{\mathrm{keV}}{\mathrm{cm}^{3}}\right)\mathcal{F}\left(T_{\star}\right)  \sqrt{\frac{m_{_{W_\R}}}{\mathrm{eV}}} \left(\frac{\mathcal{W}^{ b}_{i,e}}{10^{12} \mathrm{GeV}}\right)^{2} ,
\eeq
where $\mathcal{F}\left(T_{\star}\right) \equiv\left(g_{\star}\left(T_{\star}\right) / 3.36\right)^{3 / 4} 
( 3.91/g_{\star S}(T_{\star}))$.

The current energy density is proportional to the square of the field value 
at the end of inflation $(\mathcal{W}^{ b}_{i,e})^2$ which is obtained from Eq.~\ref{Eq:endinflation}.
Since the $\rho_{\mathrm{DM}}=1.25 \frac{\mathrm{keV}}{\mathrm{cm}^{3}}$~\cite{Planck:2018jri}, 
the relic density of the non-abelian vector can be found as follows,
\beq
\frac{\Omega_{\mathcal{W}^b}}{\Omega_{\mathrm{DM}}} \simeq 5.2 \mathcal{F}\left(T_{\star}\right)  
\sqrt{\frac{m_{_{W_\R}}}{\mathrm{eV}}} \left(\frac{\mathcal{W}^{ b}_{i,e}}{10^{12} \mathrm{GeV}}\right)^{2}.
\eeq

After considering the homogeneous field, we should care about the fluctuation analysis. 
One important constraint in the above scenario comes from the cosmic microwave background (CMB) limits on the isocurvature fluctuations. 
During inflation, adiabatic perturbations are generated.
Since the inflaton decays to other components of the universe 
during the reheating, these adiabatic fluctuations are transmitted to these product particles. 
However in our scenario, in addition to the inflaton during inflation, the dark gauge bosons exist.  
Since inflaton is not responsible for the production of the dark gauge bosons during inflation, 
another source of perturbation known as isocurvature fluctuations are generated. 
Isocurvature fluctuations may not match other particle fluctuations and 
will leave footprints on the CMB. Since no such fluctuations 
have been observed yet~\cite{Planck:2018jri}, our scenario is constrained. However, 
this constraint is mostly sensitive to $H_I$ and not model parameters.  
The isocurvature constraint in our model should be very similar to that of the dark photon model~\cite{Alonso-Alvarez:2019ixv}. 

It should be mentioned that the non-minimal coupling scenario suffers from ghost instability. 
In the short wavelength, the kinetic term of the longitudinal mode of the vector field has a wrong sign.
Thus, some alternative approaches have been introduced~\cite{Nakayama:2019rhg,Nakayama:2020rka,Cline:2003gs,Sbisa:2014pzo,Karciauskas:2010as,
Himmetoglu:2008zp,Himmetoglu:2009qi}.
For instance, in Ref.~\cite{Nakayama:2019rhg}, 
the author suggests modifying the kinetic term of the gauge boson instead of adding a non-minimal coupling to the Ricci scalar in the action. 
Ref.~\cite{Nakayama:2019rhg} introduces a coupling between the inflaton and gauge bosons in the form of $f^2(\phi) F_{\mu \nu}F^{\mu \nu}$, 
where $f(\phi)$ is a function of the inflaton field $\phi$. If $f^2(\phi)$ redshifts as $a^{2}$ or $a^{-4}$ during inflation 
while approaching $f(\phi)\to 1$ by the end of inflation, 
one can get a consistent model for a vector coherent oscillation where during inflation the vector field has homogeneous condensate 
and at the late time, it has a coherent oscillation without any instability problem in the theory.
The same model with the non-abelian vector field coupling to a scalar field via the gauge kinetic function is also feasible. 
The important phenomenological difference between the scenario with non-minimal coupling to gravity 
and the one with the modified kinetic term is the constraint coming from isocurvature perturbation.
Future measurements of $H_I$ would favor one of these scenarios.
In the following, however, we will be oblivious to the cosmological UV completion of the model and 
only study the terrestrial constraints on the model. 

%%%%%%%%%%%%%%%%DMcandidates%%%%%%%%%%%%%%%%%%
\section{DM Phenomenology}
\label{sec:Pheno}
%%%%%%%%%%%%%%%%DMcandidates%%%%%%%%%%%%%%%%%% 
Light dark non-abelian gauge boson $(W_\R^3)$ that couples to electrons has a similar phenomenology to the dark photon, 
which has been discussed extensively in the literature. In particular, the following constraints apply to $W_\R^3$.   
\begin{enumerate} 
\item Light $W_\R^3$ can be produced inside hot ($\sim O(20$ MeV)) and dense stars. 
If $W_\R^3$ is sufficiently weakly coupled, it can escape the star without re-scattering. 
Since it carries some energy, this process contributes to the cooling rate of the star. 
The bounds from the Sun, Horizontal Branch (HB), and Red Giants (RG) are shown in green, orange, and red shaded regions, respectively, 
in Fig.~\ref{Fig:bounds} ~\cite{An:2014twa,An:2013yfc,Redondo:2013lna}. 
If the dark sector's particle mass is close to the plasma frequency, the dark sector particle can be produced resonantly 
and thus leave a more significant impact on the cooling rate of the star. 
Thus, these cooling rates are more sensitive at some particular mass of $W_\R^3$. 

\begin{figure}[!h]
\centering
\includegraphics[width=12cm,height=8cm]{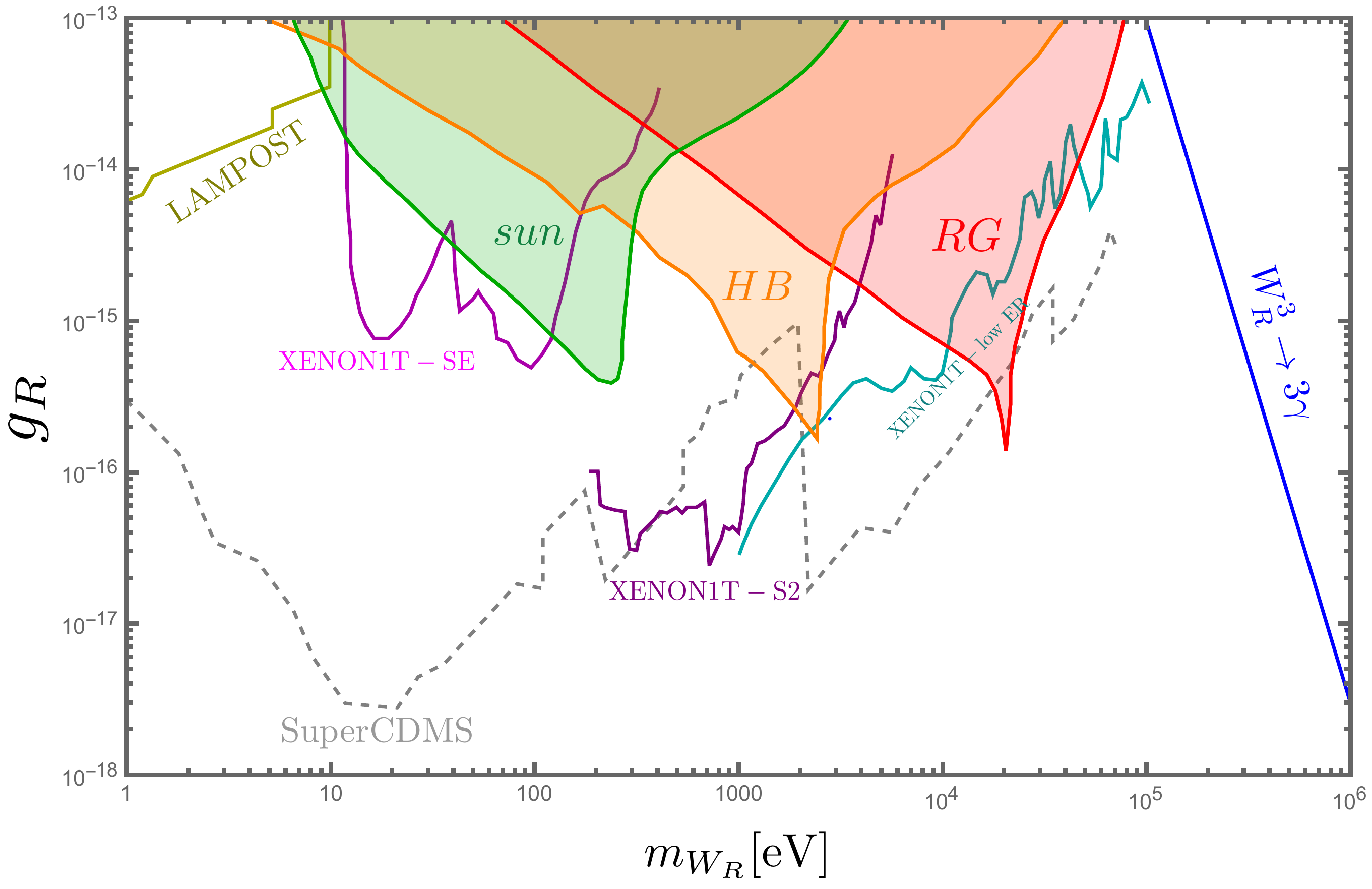}
\caption{The current bounds on $W_\R^3$ are demonstrated.  
The shaded regions are the constraints from stellar cooling: Sun (green), Horizontal Branch (orange), and Red Giants (red)
~\cite{An:2014twa, An:2013yfc, Redondo:2013lna}. 
The bound from the decay of $W_\R^3 \to 3 \gamma$ is shown by the blue line~\cite{An:2014twa}. 
Various XENON1T analyses probe different regions of parameter: the low electron recoil region 
is shown by the dark-cyan line~\cite{XENON:2020rca}, the analysis on S2 signal (as explained in the text) 
gives the purple line~\cite{XENON:2019gfn}, and the extension of this analysis in lighter mass regime provides 
a constraint shown by the magenta line~\cite{XENON:2021myl}. The projected bound coming from the absorption of 
DM in the super-CDMS experiment is shown by the dashed gray-line~\cite{Bloch:2016sjj}. 
The LAMPOST constraint is shown by the dark-yellow line~\cite{Chiles:2021gxk}.}
\label{Fig:bounds}
\end{figure}

\item  Due to the coupling of dark gauge bosons with electrons, $W_\R^3$ can decay 
to three photons through a loop of electrons.
This process has the following decay rate~\cite{Redondo:2008ec,Pospelov:2008jk,McDermott:2017qcg,Bhoonah:2018gjb}:
\begin{equation}
\Gamma_{W_\R^3 \rightarrow \gamma \gamma \gamma} =
\frac{17 \left(g_\R\right)^2 \alpha^{3}}{2^5 3^6 5^3 \pi^{2}} \frac{m_{_{W_\R}}^{9}}{m_{e}^{8}}.
\end{equation}

The decay of $W^3_R$ can alter the ionization history, and thus leave a footprint in the cosmic microwave background (CMB) data. 
However, the constraint becomes faint if $\tau_{W^3_R \to 3 \gamma} \gtrsim 10^{25} \ s$~\cite{Fradette:2014sza,Poulin:2016anj}. 
Furthermore, one can find the constraint on the energy injection of $\gamma$-rays 
coming from the decay of $W^3_R \to 3 \gamma$ in the present time~\cite{Pospelov:2008jk,An:2014twa}:
\beq
E_\gamma \frac{d \phi_{gal}}{dE_\gamma} = \frac{\Gamma_{W^3_R \to 3 \gamma}}{4 \pi m_{_{W_\R}}} E_\gamma\frac{dN}{dE_\gamma}\rho_{sol} R_{sol},
\eeq
where $\rho_{sol}$ is the energy density of $W^3_R$ the sun's position, 
and  $R_{sol} \simeq 8.3\  \text{kpc}$ is the distance between the sun and the center of the galaxy. 
A limit from the observation of the diffuse $\gamma$-ray was estimated in~\cite{Redondo:2008ec} 
using the monochromatic photon injection found in~\cite{Yuksel:2007dr}. 
Interestingly, the constraints from CMB and the galactic diffuse $\gamma$-ray are overlapping 
and thus are shown with a single blue line in Fig.~\ref{Fig:bounds}. 

\item When DM enters the XENON1T experiment, it can scatter off of xenon nuclei or an electron. 
For the case of nuclei recoil, a scintillation is generated as a result of the de-excitation of the excited xenon to its ground state.
This scintillation signal is called S1.  In the case of electron recoil, one or a few electrons may be freed via atomic ionization. 
A series of electric fields are used to drift the freed electrons from its interaction point into gaseous xenon, 
which produces a secondary scintillation signal called S2 (the purple line in Fig.~\ref{Fig:bounds})~\cite{XENON:2019gfn}. 
The ratio $S1/S2$ is used to distinguish between low electron recoil (ER) 
(dark-cyan line in Fig.~\ref{Fig:bounds})~\cite{XENON:2020rca} from nuclear recoil
\footnote{The analysis on nuclear recoil is not sensitive to the region of the parameter of our interest.}.
The Magenta line in Fig. ~\ref{Fig:bounds} is a development on XENON1T-S2 analysis, written as XENON1T-SE on the plot, 
to increase the sensitivity to smaller mass region~\cite{XENON:2021myl}. 
\footnote{It is worth mentioning that the XENON1T experiment reported an excess of the ER 
events below 7 keV with a 3-4 $\sigma$ local significance~\cite{XENON:2020rca}.
Similar to Ref.~\cite{Alonso-Alvarez:2020cdv,Okada:2020evk}, our model could also explain this excess at the benchmark of 
\beq
m_{_{W_\R}}=2.8~\mathrm{keV},\hspace{0.2 in}  g_{\R}=2.26 \times 10^{-16}. \nonumber
\eeq
However, further measurements with XENONnT ruled this excess out~\cite{XENON:2022ltv}.}
\item  The absorption of light DM by cold atoms in direct detection can also provide strong constraints on DM interactions. 
The absorption cross section is similar to that of the photon which carries the momentum of $m_{_{W_\R}}$, 
but with a different coupling to electrons~\cite{An:2014twa}
\beq
\text{Rate per atom} \simeq \frac{\rho_{DM}}{m_{_{W_\R}}} \left( \frac{g_\R}{e}\right)^2 \sigma_\gamma 
(|\vec q| = m_{_{W_\R}}),
\eeq
where $\rho_{DM}$ is the local energy density of $W^3_R$ DM. The dashed-gray line shows the projected 90\% C.L. 
sensitivities for Super-CDMS using Ge (20 kg-years)~\cite{Bloch:2016sjj}.

\item Recently, the LAMPOST ( Light $A'$ Multilayer Periodic  Optical SNSPD Target) experiment has searched for a DM 
(e.g, dark photon, axion, or in our case dark non-abelian gauge boson) that converts to a photon. 
This experiment uses a superconducting nanowire single-photon detector (SNSPD) to detect the incoming photon. 
The LAMPOST experiment is sensitive to $O(\rm{eV})$ DM~\cite{Chiles:2021gxk}. 
The region of the parameter space that the LAMPOST experiment can probe is shown by the dark-yellow line in Fig.~\ref{Fig:bounds}. 
This bound is sensitive to the kinetic mixing coupling, where we have assumed $\epsilon_W (v_\phi/\Lambda_2)^2 \simeq g_\R e/ (16 \pi^2)$. 
 
\item Since $W_R^3$ couples to $e_R$, it will contribute to electron $g-2$, through diagram~\ref{Fig:feyn}.   
The contribution of $W_R^3$ is the following~\cite{Queiroz:2014zfa,Leveille:1977rc,Ayazi:2021dca}:
\beq
\Delta a_e^{NP}  \simeq  - \frac{1}{3} \left(\frac{g_\R m_e}{ \pi m_{_{W_\R}}} \right)^2 \left(1+ O\left(\frac{m_e}{m_{e'}}\right)\right),
\eeq
Where the contribution of $W_\R^\pm$ is $O (m_e/m_{e'})$ suppressed. 
The new physics contribution to $(g-2)_e$ has to  $\Delta a_e^{NP} \lesssim 10^{-13}$~\cite{Parker:2018vye,Morel:2020dww}. 
Since we are considering very weakly coupled 
(e.g, $g_\R < 10^{-12}$) and $O(\rm{eV-MeV})$ DM, this constraint does not appear on the constraint plot.

\begin{figure}[h]
\centering
\includegraphics[width=4cm,height=4cm]{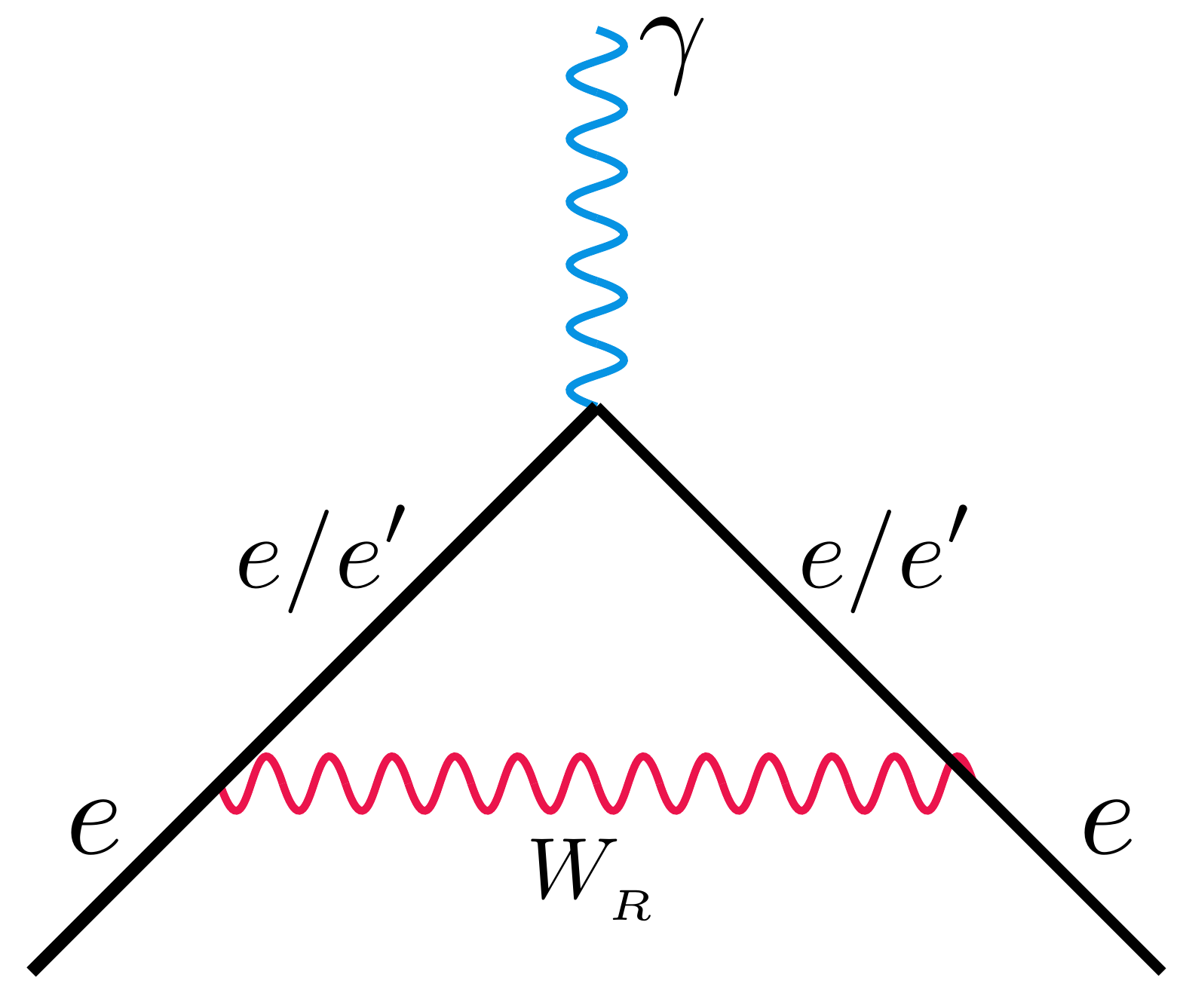}
\caption{The Feynman diagram contributes to $(g-2)_e$ that involves $W_\R$.  }
\label{Fig:feyn}
\end{figure}

\end{enumerate}

The aforementioned constraints are due to the similarities between $W_\R^3$ and dark photon. 
However, due to the non-abelian nature of $\SU$, we need to consider the constraint coming 
from the self-interaction (SI) between dark gauge bosons. 
The SI between DM is under great debate. On the one hand, 
large self-interaction can explain the small scale structure problems 
that are associated with CCDM\footnote{collision-less cold dark matter}(Ref. \cite{Tulin:2017ara}, and references therein) . On the other hand, large self interaction of DM (e.g, $ \sigma/ m_{\DM} > 0.47 \  \text{cm}^2/\text{g}= 4.59 \times 10^{3} \gev^{-3}$ at $ v_{\DM} \simeq 10^{-3}$) is excluded by the observation of merging clusters~\cite{Harvey:2015hha,Kaplinghat:2015aga}. 
The elastic self-interaction between dark gauge bosons is the following:
\beq
\sigma_{(W_R W_R \to W_R W_R)} = \frac{1}{2}\int dt \frac{t}{2 \left(\frac{s}{2} - m_{_{W_\R}}^2\right)^2} \frac{1}{64 \pi s}|\cM|^2,
\eeq
where $|\cM|^2$ is the squared matrix element of the $W_R$ self-interaction, and $s,t$ are the the ordinary Mandelstam variables. In the non-relativistic limit, this cross section is 
\begin{align}
\sigma_{(W_R W_R \to W_R W_R)} &=\frac{g_\R^4 (320 m_{_{W_\R}}^8 - 1200 m_{_{W_\R}}^6 m_\phi^2 + 8801 m_{_{W_\R}}^4 m_\phi^4 -4208 m_{_{W_\R}}^2 m_\phi^6 + 520 m_\phi^8)}{1152 \pi m_{_{W_\R}}^2 m_\phi^4( m_\phi^2 - 4 m_{_{W_\R}}^2)^2}\nonumber\\
& \simeq \frac{65}{144 \pi} \frac{g_\R^4}{m_{_{W_\R}}^2} \hspace{0.5 in} \text{for} \hspace{0.5 in}m_\phi \gg m_{_{W_\R}}.
\end{align}
Therefore, for $m_{_{W_\R}} \lesssim  10^7  \text{eV} \times g_\R^{4/3}$, 
merging cluster does not impose any constraint.

One may also wonder about the phenomenology of $W_\R$s at colliders.  
With high luminosity LHC (HL-LHC), the minimum cross section 
that can be probed is roughly $\sigma^{\min}_{NP} \simeq O(1)\  \text{ab}$, 
since the integrated luminosity is expected to be about $3\ \text{ab}^{-1}$~\cite{CMS:2013xfa}. 
The cross section for producing $W_\R$ at colliders is proportional to $ \left(g_\R\right)^2$. 
For $g_\R < 10^{-12}$, the production cross section will be orders of magnitude smaller 
than $\sigma^{\min}_{NP}$. Thus, $W_\R$ is unlikely to be produced at the upgraded LHC.

\subsection{$e'$ Phenomenology}

\begin{figure}[!h]
\centering
\includegraphics[width=6cm,height=3.5cm]{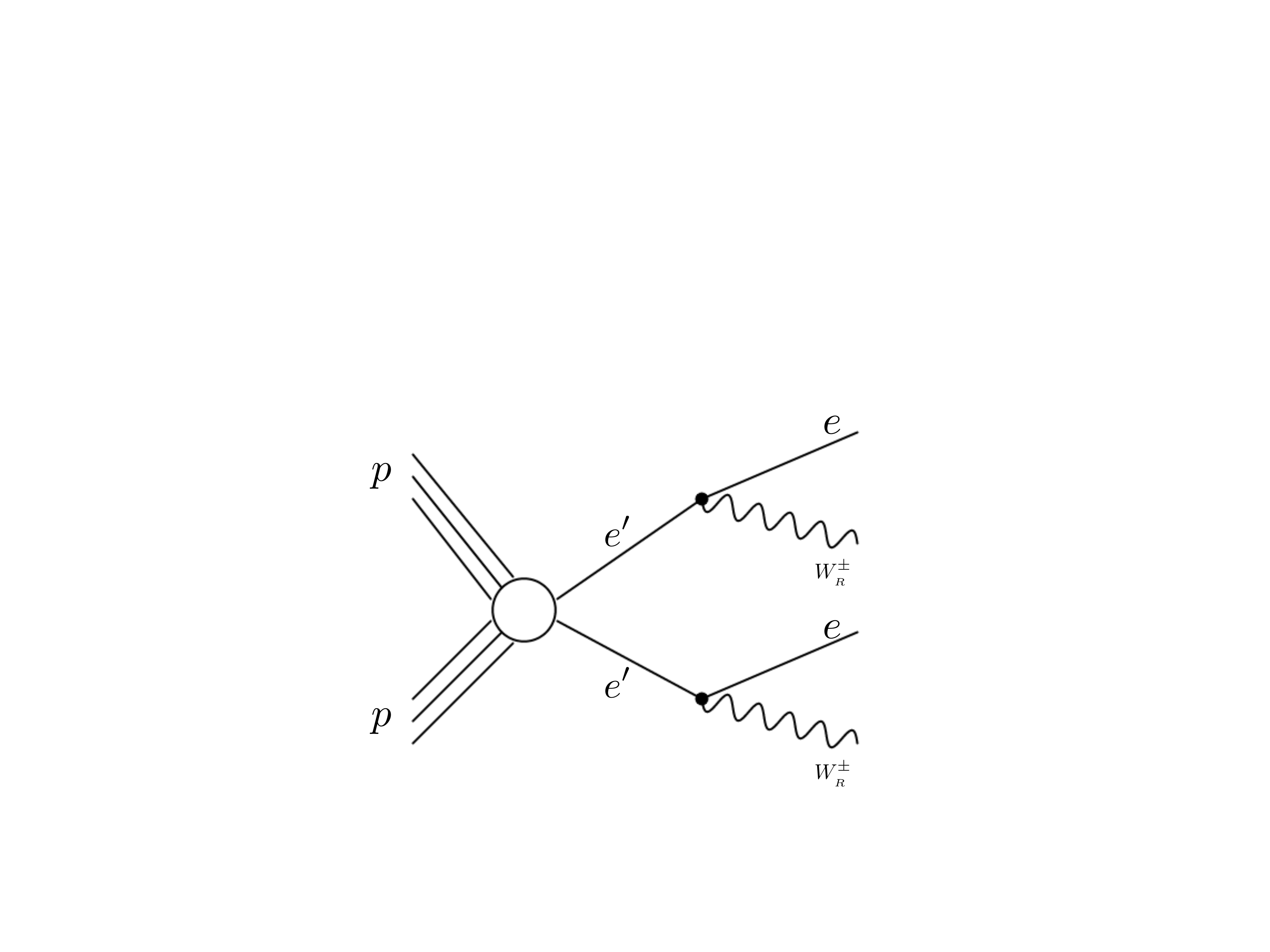}
\caption{ The Feynman diagram for $e'$ pair production at the LHC.}
\label{Fig:eprim}
\end{figure}

The production of the $e'$ at colliders is proportional to $ g_Y^2\simeq O(1)$. 
Therefore, depending on the mass of $e'$, its production at LHC or FCC is plausible ~\cite{FCC:2018evy}. 
The pair production of $e'$ can happen through the following s-channel process 
at the LHC where both of $e'$s decay to $e W_\R^\pm$:
\begin{equation}
p p \rightarrow \gamma^{*}/ Z^{*} \rightarrow e' e' \rightarrow (e W_\R^\pm) (e W_\R^\pm). \nonumber
\end{equation}
Since we assume small gauge coupling, the $e'$ should be a long-lived particle.
However, because $e'$s have electromagnetic charge, they will leave two tracks in the tracker, 
where the curve of the tracks is dependent on the mass of $e'$. 
As a result, the detector signature is two displaced vertices from the primary $p p$ interaction point. 
In the final state, we have two displaced, isolated leptons (two electrons) and missing energy due to the $W_\R^\pm$s. 
The related Feynman diagram is shown in Fig.~\ref{Fig:eprim}.
Recently, the CMS collaboration has published their results on such searches for the center of mass-energy $\sqrt{s}=13$~TeV with an integrated luminosity of $118$ fb$^{-1}$.
The current bound corresponds to $m_{e'} > 610\ \text{GeV}$~\cite{CMS:2021kdm} for $e'$ with a life time $c \tau_{e'} < 1$ cm. 
This bound on the $e'$ mass, ensures us the stability of the $W_\R^{\pm}$ since the only decay channel for dark gauge boson  $(W_\R^{\pm}\rightarrow e e')$ is kinematically forbidden.

%%%%%%%%%%%%%%%%DMcandidates%%%%%%%%%%%%%%%%%%
\section{Summary}
\label{sec:Sum}
%%%%%%%%%%%%%%%%DMcandidates%%%%%%%%%%%%%%%%%% 

In this paper, we examined a multi-component dark matter (DM) scenario with light-feebly interacting non-abelian dark gauge bosons.
In this region of the parameter space, $W_\R^3$ and $W_\R^\pm \equiv W_\R^1 \pm i W_\R^2$
associated with the gauge $SU(2)_{\R}$ are the DM candidates. 
Due to their small gauge coupling, the evolution of $W_\R$s in the early universe is non-trivial.  

We charged right-handed electrons under the $\SU$ symmetry to provide a portal between the dark and the standard model sectors. 
We investigated the production of $W_\R$ dark gauge bosons via freeze-in and vector misalignment mechanisms.
We found that for $g_\R \simeq 10^{-12}$, the relic density of $W^3_\R$ freezing in from $e e \to \gamma W_\R^3$ 
can account for the total DM abundance in the universe. However, the obtained relic abundance for $W_\R^\pm$ via 
the freeze-in mechanism is negligible.

For smaller gauge couplings, the vector misalignment mechanism can be used
to produce the non-abelian dark gauge bosons.
We extended the action of the non-abelian vector field with non-minimal coupling to the Ricci scalar term
and showed the obtained equation of motion (EOM)
for the spatial component of a homogeneous non-abelian vector field is the same 
as the harmonic oscillator with Hubble damping term.
The non-minimal coupling to Ricci scalar leads to an effective mass in the EOM,
which has the most important effect on the inflationary era.
In contrast to the $U(1)$ Stuckelberg mechanism, 
we must consider massless gauge bosons before $\SU$ is spontaneously broken by a scalar $\phi$. 
However, as long as the $\SU$ breaking scale is before the Hubble rate matches the $W_\R$ mass, 
the story of non-abelian gauge symmetries becomes similar to dark photons. 
In the late epoch when the Hubble parameter is much smaller than the mass of the non-abelian vector, the field oscillates 
around the minimum of the potential. This coherent oscillation acts as a non-relativistic matter 
and its abundance can explain the observed relic density of DM in the present time.
We should mention that the vector misalignment mechanism can produce all components of the 
light dark gauge boson.

Due to the interaction between $W_\R$ and right-handed electrons, 
this model can be probed in various astrophysical observations and terrestrial experiments.  
Particularly, we demonstrated the parameter space excluded by the stellar cooling, 
the CMB and the galactic diffuse $\gamma$-ray observations, and different direct detection experiments 
(like, XENON1T, Super-CDMS, and LAMPOST). 
%Moreover, we determined the parameter values
%which can explain the reported electron recoil excess in the XENON1T experiment. 
Other bounds such as DM self-interaction or electron $g-2$ provide mild constraints on this model. 

Therefore, the phenomenology of this model is very similar to an ordinary dark photon dark matter model. 
The existence of a heavy electron-like particle and $W^\pm_\R$ distinguishes this model from dark photon models. 

%%%%%%%%%%%%%%%%%%%%%
\section*{Acknowledgments}
\label{sec:ack}
%%%%%%%%%%%%%%%%%%%%%
We would like to thank N. Khosravi, H. Mehrabpour and P. Schwaller for useful discussions. 
The work of FE was supported by the Cluster of Excellence Precision Physics, 
Fundamental Interactions, and Structure of Matter (PRISMA+ EXC 2118/1) funded 
by the German Research Foundation(DFG) within the German Excellence Strategy 
(Project ID 39083149), and by grant 05H18UMCA1 of the German Federal Ministry 
for Education and Research (BMBF).
%%%%%%%%%%%%%%%%%%%%%%

\appendix 
\numberwithin{equation}{section}
\section{Sub-dominant Freeze-In Processes }
\label{app:FIsmall}

If the operator responsible for the production of $W_\R$ is renormalizable, Ref.~\cite{Hall:2009bx} shows that the production of $W_\R$ is efficient at lower temperatures when $T \sim m_{_{W_\R}}$. 
If the operator is non-renormalizable, however, the production of $W_\R$ is dominated at high temperatures $T \simeq T_{\rm {Max}}$~\cite{Elahi:2014fsa,Garcia:2017tuj,Bernal:2019mhf}, where $ T_{\rm {Max}} = T_{\rm{RH}}$ assuming the inflaton decays instantly. 
Let us denote $T_\phi$ to indicate the temperature at which $\phi$ acquires a vev. In the following, we compute the relic abundance of $W_\R$ in the two regimes of $T> T_\phi$ and $T < T_\phi$. We will see that the processes discussed in this appendix lead to a negligible abundance of $W_\R$.

%%%%%%%%%%%%%%%%%%%%%%%%%%%%%%%%
\subsection{$T_\phi < T < T_{\rm{RH}} < \Lambda_2$}
\begin{figure}[t]
\centering
\includegraphics[width=7cm,height=3cm]{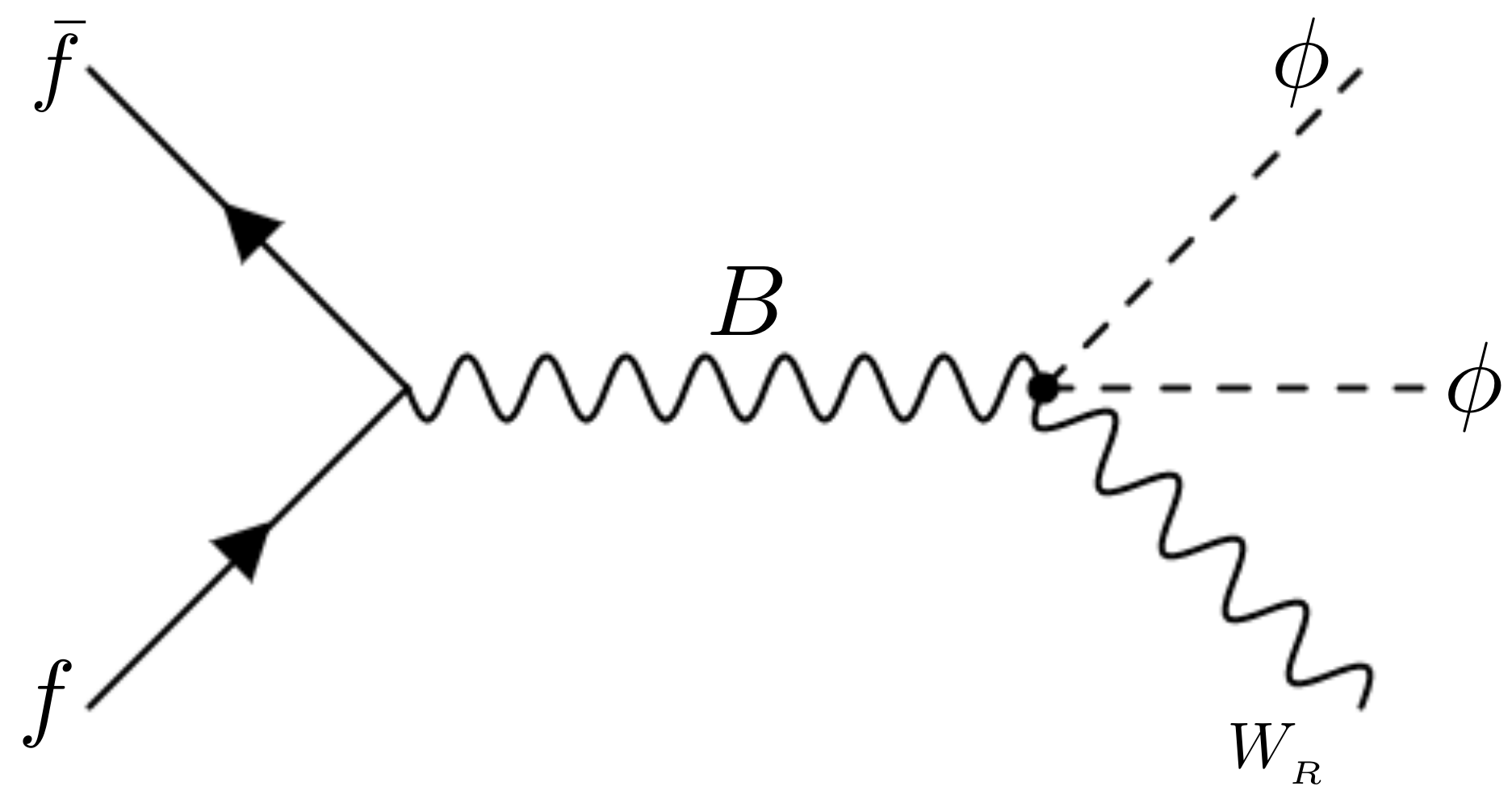}
\caption{The freeze-in production of $W_\R$ in the early universe before $\phi$ acquires a vev via the non-renormalizable operator $ \frac{\phi^\dagger \tau^a \phi}{\Lambda_2^2} W_{\R,\mu \nu}^{a} B^{\mu \nu}$. The relic abundance of $W_\R$ coming from this process depends on the UV parameters. However, a realistic estimate reveals that the amount of $W_\R$ produced through this process is orders of magnitude smaller than the observed relic density of dark matter. }
\label{Fig:FeynamFI1}
\end{figure}

%%%%%%%%%%%%%%%%%%%%%%%%%%%%%%%%%
 For $ T > T_\phi$, the non-renormalizable kinetic mixing term can contribute to the production of $W_\R$ at high temperatures via $f \bar f \rightarrow W_\R~\phi \phi$ (Fig.~\ref{Fig:FeynamFI1}), where $f$ is any fermion that has a non-zero charge under the hypercharge symmetry\footnote{SM fermions plus $e'$ and other fermions in the UV needed to cancel the anomaly of the theory.}. The Boltzmann equation of $W_\R$ in this regime is~\cite{Elahi:2014fsa}: 
 \beq
 \dot n_{W_\R}+ 3 H n_{W_\R} = \frac{T}{(4\pi)^7} \int_0^\infty ds \ s^{3/2} |\mathcal{M}|^2 K_1\left(\frac{\sqrt{s}}{T}\right) \simeq \frac{3N_f}{16 \pi^5}\frac{\epsilon_W^2 g_\R^2 \alpha_Y^2 T^8}{\Lambda_2^4},
 \eeq 
where $|\mathcal{M}|^2 \simeq 16 \pi^2 \epsilon_W^2 g_\R^2 \alpha_Y^2 s /\Lambda_2^4$, with $\alpha_Y = g_Y^2/(4\pi)$, and $N_f$ is the number of fermions with a hypercharge charge. Changing the variable to $Y_{_{W_\R}}$, we get 
\beq
\frac{dY_{_{W_\R}}}{dT} \simeq - \frac{1}{s(T)H(T)}\frac{3N_f}{16 \pi^5}\frac{\epsilon_W^2 g_\R^2 \alpha_Y^2 T^7}{\Lambda_2^4}, 
\label{eq:dYdT}
\eeq
integrating Eq.~\ref{eq:dYdT}, in the limit of $([ T_\phi, T_{\rm{RH}}])$, the yields of the $W_\R$ is obtained:
\beq
Y_{_{W_\R}} \simeq \frac{ 45 N_f  \epsilon_W^2 \alpha^2 \ g_\R^2 }{32 \pi^7 \times1.66 \times g_\star^{3/2} } \frac{M_{\text{pl}} ( T^3_{\text{RH}} - T^3_\phi)}{\Lambda_2^4}.
\label{eq:YUV}
\eeq
In this temperature range, other contributions to $W_\R$ relic abundance will be negligible as these processes are dominant at $T \sim m_{_{W_\R}}$.

As Eq.~\ref{eq:YUV} shows, the relic abundance of $W_\R$ before $\phi$ acquires a vev is sensitive to the UV scale parameters, which are irrelevant at low energy phenomenology.
Nonetheless, for illustrative purposes, let us choose the following benchmark to gain an intuition about the contribution of UV-FI to $W_\R$ production:
\begin{align*}
\Lambda_2 = 10^{7} \ \text{GeV}, \hspace{0.3 in} &T_{\text{RH}} =10^6 \ \text{GeV}, \hspace{0.3 in} T_{\phi} = 10^5\  \text{GeV}, \\
g_\R = 10^{-6} , \hspace{0.3 in}  \epsilon_W= &g_\R g_Y/(16 \pi^2) \simeq 2 \times 10^{-8}, \hspace{0.3 in}  N_f = 30. 
\end{align*}
For this benchmark, the relic abundance is 
$\Omega_{_{W_\R}} h^2 \simeq 10^{-18} \left( \frac{m_{_{W_\R}}}{\text{GeV}}\right)\ll  \Omega_{\text{DM}} h^2$. 
Hence, we can safely neglect the contribution from UV-FI to the $W_\R$ relic abundance.
  
One may also wonder about the production of $W^\pm_\R$ through the plasmon effect~\cite{Dvorkin:2019zdi}: $\gamma^\star \to W_\R^3 \to W_\R^+ W_\R^-$. However, this process is proportional to $\epsilon_W^2 g_\R^4$. 
Given that $\epsilon_W < g_\R$, this process is highly suppressed and thus will not leave any impact on CMB.  

%%%%%%%%%%%%%%%%%%%%%%%%%%%%%%%%
\subsection{$m_{_{W_\R}}< T < T_\phi $}
\begin{figure}[t]
\centering
\includegraphics[width=14cm,height=3.5cm]{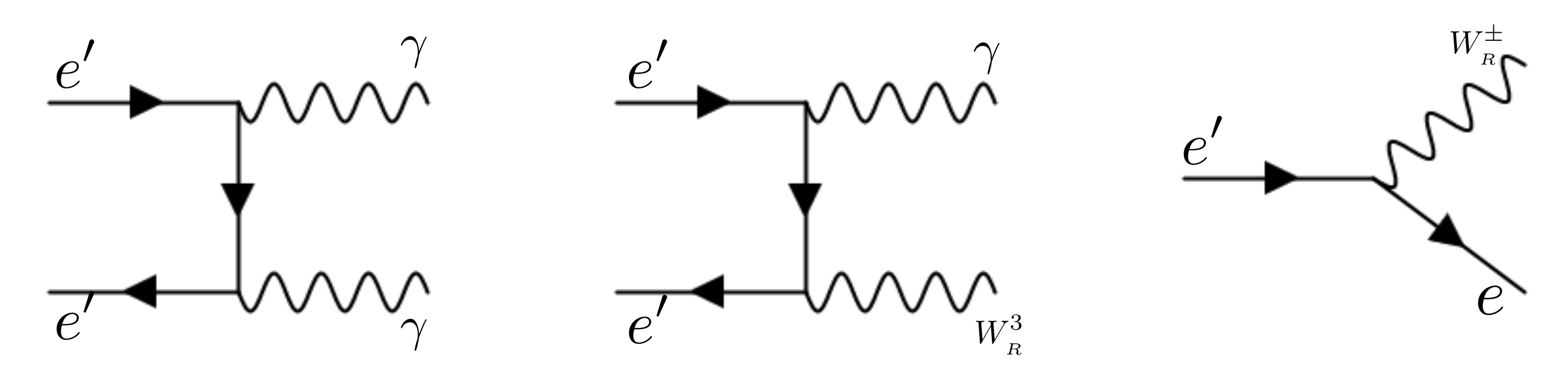}
\caption{The Feynman diagrams that influence $n_{e'}$ in the early universe. Once $e'$ becomes non-relativistic, its number density depletes rapidly because of its annihilation to a pair of photon (or B). In the limit where $g_\R \ll \alpha$, the production of $W_\R^{3,\pm}$ is very inefficient.   }
\label{Fig:FeynamFI2}
\end{figure}

%%%%%%%%%%%%%%%%%%%%%%%%%%%%%%%%%
At low temperatures $T < T_\phi, m_\phi$, the dominant diagrams leading to production of $W_\R$ are shown in Fig~\ref{Fig:FeynamFI2}. 
%In the freeze-in mechanism, the dark sector never reaches equilibrium. 
%Therefore, only the forward direction of processes ($\bar{e}~ e (\bar{e'}~ e' ) \rightarrow W^{3}_\R~\gamma~(a)$ and $e' \rightarrow e~W^{\pm}_\R~(b)$) occur, until the thermal bath does not have enough energy to produce $W^{3}_\R$ and $W^{\pm}_\R$ bosons.
After $\phi$ acquires a vev, $e'$ gains a mass  $m_{e'} \sim v_\phi$, and thus immediately becomes non-relativistic. 
In this regime, $e'$ number density decreases because of (a) annihilation to a pair of photon\footnote{ For simplicity, in this subsection we use $\gamma$ instead of B.}, (b) annihilation to $W_\R^3 \gamma$, or (c) decay to $e W^\pm_\R$:
\beq
\dot n_{e'}+ 3 H n_{e'} = - \langle \sigma v \rangle_{_{e' e' \to \gamma \gamma}} (n_{e'}^2 - n_{e',\text{eq}}^2) -   \langle \sigma v \rangle_{_{e' e' \to W^3_\R \gamma}} n_{e'}^2 - n_{e'} \Gamma_{e' \to e W^\pm_\R}.
\eeq
In the regime where $g_\R \ll \alpha$, the production of $W_\R$ from $e'$ is overwhelmed by $e' e' \to \gamma \gamma$. For $g_\R \lesssim 10^{-10}$, the relic abundance of $W_\R$ from $e'$ is roughly $\Omega_{W_\R} = \Omega_{W^\pm_\R}\lesssim 10^{-26} \Omega_{\text{DM}}$.  This relic abundance is estimated by assuming that $e'$ freezes out after its annihilation to a pair of photon, and the remaining $e'$ decays to $e W^\pm_\R$ through the dark gauge coupling. 

\section{Equation of Motion for a Non-Abelian Vector Boson}
\label{app:EOM}

The action for the non-abelian vector field with non-minimal coupling to the Ricci scalar is given by,

\beq
S=\int \mathrm{d}^{4} x \sqrt{-g}\left(\frac{1}{2}\left(M_{\mathrm{pl}}^{2}+
\frac{\kappa}{6} W_{_{R} \mu}^b W_{_{R} }^{b~\mu}\right) R-
\frac{1}{4} W^{b}_{_{R}\mu\nu} W^{ b~\mu\nu}_{_R}-
\frac{1}{2} m_{_{W_\R}}^{2} W_{_{R}\mu}^b W_{_{R} }^{b ~\mu}\right).
\eeq

Accordingly, the equation of motion (EOM) is obtained as follows,
\beq
\partial_{\mu}\left(\sqrt{-g}  W^{ b~\mu\nu}_{_R} \right)=\sqrt{-g} (\frac{\kappa}{6} R - 
m_{_{W_\R}}^{2} ) W_{_{R} }^{b~\nu}. 
\eeq 

If we consider a spatially homogeneous field, we can neglect the spatial derivatives. Using the FRW 
metric, $\sqrt{-g} =a^3$ which $a$ is the scalar factor,
\beq
\partial_{0}[a^3 W^{ b~0\nu}_{_R}]= a^3 (\frac{\kappa}{6} R - m_{_{W_\R}}^{2} ) W_{_{R} }^{b~\nu}. 
\eeq

For the time component $\nu=0$,

\beq
3 H W^{b}_{_{R} 00} + \partial_{0} W^{b}_{_{R} 00} = - (\frac{\kappa}{6} R - m_{_{W_\R}}^{2} ) W_{_{R} 0}^b,
\eeq

where $H=a/\dot{a}$ is the Hubble parameter. Since
\beq
W^{b}_{_{R} 00}= \partial_{0} W_{_{R} 0}^b -\partial_{0} W_{_{R} 0}^b + g_{R} f^b_{cd} W_{_{R} 0}^c W_{_{R} 0}^d=0,
\eeq
where $f^b_{cd}$ is the antisymmetric structure constant of the $\SU$.
As a result, $W_{_{R} 0}^b=0$.
 
For the spatial component  $\nu=i$,
\beq
 \partial_{0} W^{b}_{_{R} 0i} + H W^{b}_{_{R} 0i} +(m_{_{W_\R}}^{2} - \frac{\kappa}{6} R  ) W_{_{R} i}^b=0.
\eeq 

If we neglect the spatial derivatives and consider $W_{_{R} 0}^b=0$, the EOM for the spatial component
be obtained as follows,

\beq
\ddot{W_{_{R} i}^b}+ H \dot{W_{_{R} i}^b}+\left(m_{_{W_\R}}^{2}-\frac{\kappa}{6} R\right) W_{_{R} i}^b=0. 
\eeq	

By field redefinition $\mathcal{W}^{ b}_{i} \equiv W_{_{R} i}^b/ a$ and considering 
$R= 6(\frac{\ddot a}{a}+\frac{\dot{a}^2}{a^2})=6\dot{H}+12H^2$,

\beq
\ddot{\mathcal{W}^{ b}_{i}}+3 H \dot{\mathcal{W}^{ b}_{i}}+\left(m_{_{W_\R}}^{2}+\frac{1-\kappa}{6} R\right) \mathcal{W}^{ b}_{i}=0, 
\label{Eq:EOM}
\eeq	 
 
Therefore, the EOM for the spatial component of a homogeneous non-abelian vector field is the same 
as the damped harmonic oscillator.

\section{Solving the EOM in the different eras}
\label{app:Evolution}
We can solve the EOM of a non-abelian vector field (\ref{Eq:EOM}) for the evolving universe,
\beq
\mathcal{W}^{ b}_{i}(t)=\alpha_1 e^{i \omega_- t}+\alpha_2 e^{i \omega_+ t},
\eeq
where the frequencies $\omega_{\pm}=\frac{3i H}{2} \pm \frac{1}{2} \sqrt{-9 H^2 +4(m_{_{W_\R}}^{2}+\frac{1-\kappa}{6} R)}$.
We can determine the above equations in different eras in the evolving universe.

%%%%%%%%%%%%%%%
\subsection{Inflation era}
%%%%%%%%%%%%
During inflation, $m_{_{W_\R}}$ is zero since the SSB does not happen yet. The Hubble parameter 
is constant $H_{I}$ and $R=12 H^2_{I}$, so we have the damped harmonic oscillator with the constant frequency~\cite{Alonso-Alvarez:2019ixv},
\beq
\omega_{\pm}=\frac{i H_I}{2} \beta_{\pm},
\eeq
where $\beta_{\pm}= 3 \pm  \sqrt{1+8 \kappa}$.

\beq
\mathcal{W}^{ b}_{i}(t) =\mathcal{W}^{ b}_{i,s}\left(c_{1} \mathrm{e}^{-\frac{1}{2} \beta_{-} H_{I} t}+c_{2} \mathrm{e}^{-\frac{1}{2} \beta_{+} H_{I} t}\right),
\eeq
where the $\mathcal{W}^{ b}_{i,s}$ is the initial value of the field at the start of inflation. For $ \kappa > 1$, to avoid a trans-Planckian field excursion, $\beta_+$ is positive and $\beta_-$ is negative, so after passing enough time 
the term is proportional $\mathrm{e}^{-\frac{1}{2} \beta_{-} H_{I} t}$  will dominant. So we have,
\beq
\mathcal{W}^{ b}_{i}(t) \simeq \mathcal{W}^{ b}_{i,s} c_{1} \mathrm{e}^{-\frac{1}{2} \beta_{-} H_{I} t}.
\eeq
Since during inflation, the scale factor has a relation with the Hubble parameter ($a= e^{H_{I}t}$),
\beq
\mathcal{W}^{ b}_{i}(a) \simeq \mathcal{W}^{ b}_{i,s} (\frac{a}{a_s})^{-\frac{1}{2}\beta_-}.
\eeq
We can use the number of e-folds,
\beq
\mathcal{W}^{ b}_{i}(a) \simeq \mathcal{W}^{ b}_{i,s}  \mathrm{e}^{-\frac{1}{2} \beta_{-} N(a)}.
\eeq
So the value of the field at end of inflation can express as a function of the total number of e-folds of inflation,
\beq
\mathcal{W}^{ b}_{i,e} \simeq \mathcal{W}^{ b}_{i,s}  \mathrm{e}^{-\frac{1}{2} \beta_{-} N_{tot}}.
\eeq

%%%%%%%%%%%%%%%%%%%
\subsection{Post-Inflation era}
%%%%%%%%%%%%%%%%%%%

1. In the first part the non-abelian gauge field has zero mass since the spontaneous 
symmetry breaking (SSB) happens in the middle of the radiation era, so $m_{_{W_\R}}=0$. 
Also, in the radiation era $R=0$.
As result, the frequencies become $\omega_{+}=3iH$ and 
$\omega_{-}=0$ and we have the over-damped harmonic oscillator,
\beq
\mathcal{W}^{ b}_{i}(t)=\alpha_{1} +\alpha_{2} \mathrm{e}^{-3Ht}.
\eeq
After enough time evolution, the first term will be dominant and the value of the field will be constant.
Indeed this constant value is the value of the field at the end of inflation,
\beq
\mathcal{W}^{ b}_{i}(t) \simeq \mathcal{W}^{ b}_{i,e} .
\eeq
The vector field has a vanishing time derivative.
\\
2. After the SSB, the vector boson acquires mass. At this point $m_{_{W_\R}} \sim H_{\star}$ and the evolution of the 
vector field switch from exponential decay (constant value) to an oscillation mode. 
In another word, the field rolls down the potential and starts to oscillate.
We can consider the oscillation modes of the vector field as DM degrees of freedom.
Here, $H_{\star}$ denotes the Hubble parameter at the DM production time.
\\
3. For the late universe and  in the limit $m_{_{W_\R}} \gg H $, the frequencies become
$\omega_{\pm}=\frac{3i H}{2} \pm m_{_{W_\R}}$,
\beq
\mathcal{W}^{ b}_{i}(t)=\gamma_{1} \mathrm{e}^{-3Ht/2}~\mathrm{e}^{-i m_{_{W_\R}} t} +\gamma_{2} \mathrm{e}^{-3Ht/2}~\mathrm{e}^{i m_{_{W_\R}} t}.
\eeq
We can find the above constants, by matching to the time of the DM production, $t_{\star}$, 
\beq
\mathcal{W}^{ b}_{i}(t)= \mathcal{W}^{ b}_{i,e} (\frac{a_{\star}}{a})^{\frac{3}{2}} \cos(m_{_{W_\R}} (t-t_{\star})),
\eeq
where $a_{\star}$ is the scale factor at the DM production time.

\bibliographystyle{JHEP}
\bibliography{Light_NA_GB}

\end{document}